\documentclass[nofootinbib,prd]{revtex4}%
\usepackage{amsmath}
\usepackage{amsfonts}
\usepackage{amssymb}
\usepackage{graphicx}%
\begin{document}
\title{Naked Singularities are not Singular in Distorted Gravity}
\author{Remo Garattini}
\email{Remo.Garattini@unibg.it}
\affiliation{Universit\`{a} degli Studi di Bergamo, Facolt\`{a} di Ingegneria,}
\affiliation{Viale Marconi 5, 24044 Dalmine (Bergamo) Italy}
\affiliation{I.N.F.N. - sezione di Milano, Milan, Italy.}
\author{Barun Majumder}
\email{barunbasanta@iitgn.ac.in}
\affiliation{Indian Institute of Technology Gandhinagar }
\affiliation{Ahmedabad, Gujarat 382424, India}

\begin{abstract}
We compute the Zero Point Energy (ZPE) induced by a naked singularity with the
help of a reformulation of the Wheeler-DeWitt equation. A variational approach
is used for the calculation with Gaussian Trial Wave Functionals. The one loop
contribution of the graviton to the ZPE is extracted keeping under control the
UltraViolet divergences by means of a distorted gravitational field. Two
examples of distortion are taken under consideration: Gravity's Rainbow and
Noncommutative Geometry. Surprisingly, we find that the ZPE is no more
singular when we approach the singularity.

\end{abstract}
\maketitle

\section{Introduction}

Black holes are amazing astrophysical objects which are supposed to form as a
consequence of a gravitational collapse of some matter field. An important
black hole feature is the formation of a horizon preventing a far observer to
see its own singularity. The simplest non-rotating and uncharged black hole
can be represented by the Schwarzschild metric%
\begin{equation}
ds^{2}=-\left(  1-\frac{2MG}{r}\right)  dt^{2}+\frac{dr^{2}}{1-\frac{2MG}{r}%
}+r^{2}\left(  d\theta^{2}+\sin^{2}\theta d\phi^{2}\right)  ,\label{S}%
\end{equation}
which is obtained by solving Einstein's Field Equations in vacuum. As one can
easily see the horizon is located at $r_{S}=2MG$. It is immediate to recognize
that replacing $M$ with $-M$, one obtains another solution of Einstein's Field
Equations, but with a completely different structure: the singularity is no
more protected by an event horizon and it is naked\cite{Bondi}. In 1969,
Penrose suggested that there might be a sort of \textquotedblleft cosmic
censor\textquotedblright\ that forbids naked singularities from
forming\cite{Penrose}, namely singularities that are visible to distant
observers. An immediate consequence of a negative Schwarzschild mass is that
if one were to place two bodies initially at rest, one with a negative mass
and the other with a positive mass, both would accelerate in the same
direction going from the negative mass to the positive one. Furthermore if the
two masses are of the same magnitude, they will uniformly accelerate forever.
This means that a problem of classical stability emerges in such a
geometry\cite{GHI,GD}. If, from one side, naked singularities are well
examined from the classical point of view, it is non-trivial extracting
information from the quantum point of view. Nevertheless, an interesting
calculation is represented by the determination of Zero Point Energy (ZPE). It
is important to remark that usually any attempt to perform a ZPE calculation
inevitably faces UV divergences and therefore a regularization scheme is
needed. One possible way to take under control such divergences is given by a
zeta regularization. After the regularization a renormalization scheme can be
adopted\cite{RemoNS}. However, one can observe that at very high energies, it
is likely that space time itself can be distorted by quantum fluctuations. The
hope is that the distorted space time is also able to take under control the
UV divergences. To this purpose, we will explore two proposals:
\textit{Gravity's Rainbow} and Noncommutative geometry. When Gravity's Rainbow
is taken under consideration, spacetime is endowed with two arbitrary
functions $g_{1}\left(  E/E_{P}\right)  $ and $g_{2}\left(  E/E_{P}\right)  $
having the following properties%
\begin{equation}
\lim_{E/E_{P}\rightarrow0}g_{1}\left(  E/E_{P}\right)  =1\qquad\text{and}%
\qquad\lim_{E/E_{P}\rightarrow0}g_{2}\left(  E/E_{P}\right)  =1.\label{lim}%
\end{equation}
$g_{1}\left(  E/E_{P}\right)  $ and $g_{2}\left(  E/E_{P}\right)  $ appear
into the solutions of the modified Einstein's Field Equations\cite{MagSmo}%
\begin{equation}
G_{\mu\nu}\left(  E/E_{P}\right)  =8\pi G\left(  E/E_{P}\right)  T_{\mu\nu
}\left(  E/E_{P}\right)  +g_{\mu\nu}\Lambda\left(  E/E_{P}\right)
,\label{Gmn}%
\end{equation}
where $G\left(  E/E_{P}\right)  $ is an energy dependent Newton's constant,
defined so that $G\left(  0\right)  $ is the low-energy Newton's constant and
$\Lambda\left(  E/E_{P}\right)  $ is an energy dependent cosmological
constant. Usually $E$ is the energy associated with the particles deforming
the spacetime geometry. Since the scale of deformation involved is the Planck
scale, it is likely that spacetime itself fluctuates in such a way to produce
a ZPE. However the deformed Einstein's gravity has only one particle
available: the graviton. Therefore the particle probing the spacetime will be
the graviton produced by the fluctuations of the spacetime itself. Note that
the Rainbow's functions distort in different ways depending on the background.
For example, for the Schwarzschild background one gets%
\begin{equation}
ds^{2}=-\left(  1-\frac{2MG\left(  0\right)  }{r}\right)  \frac{dt^{2}}%
{g_{1}^{2}\left(  E/E_{P}\right)  }+\frac{dr^{2}}{\left(  1-\frac{2MG\left(
0\right)  }{r}\right)  g_{2}^{2}\left(  E/E_{P}\right)  }+\frac{r^{2}}%
{g_{2}^{2}\left(  E/E_{P}\right)  }\left(  d\theta^{2}+\sin^{2}\theta
d\phi^{2}\right)  ,\label{Sline}%
\end{equation}
where $G\left(  0\right)  $ is the low-energy Newton's constant. For the
Friedmann-Lema\^{\i}tre-Robertson-Walker (FLRW) metric describing a
homogeneous, isotropic and closed universe with line element, the Rainbow's
Gravity distortion becomes\cite{RGMS,BFRW,BM1}%
\begin{equation}
ds^{2}=-\frac{N^{2}\left(  t\right)  }{g_{1}^{2}\left(  E/E_{P}\right)
}dt^{2}+\frac{a^{2}\left(  t\right)  }{g_{2}^{2}\left(  E/E_{P}\right)
}d\Omega_{3}^{2}~,\label{FRWMod}%
\end{equation}
where $N=N(t)$ is the lapse function taken to be homogeneous and $a(t)$
denotes the scale factor. Fixing our attention on the static case, we
generalize the line element $\left(  \ref{Sline}\right)  $ in the following
way%
\begin{equation}
ds^{2}=-\frac{N^{2}\left(  r\right)  }{g_{1}^{2}\left(  E/E_{P}\right)
}dt^{2}+\frac{dr^{2}}{\left(  1-\frac{b\left(  r\right)  }{r}\right)
g_{2}^{2}\left(  E/E_{P}\right)  }+\frac{r^{2}}{g_{2}^{2}\left(
E/E_{P}\right)  }\left(  d\theta^{2}+\sin^{2}\theta d\phi^{2}\right)
,\label{LineGRw}%
\end{equation}
where $N=N\left(  r\right)  $ is the lapse function and $b\left(  r\right)  $
is termed as the shape function and its range of variability depends on case
to case. For example for the Schwarzschild metric we have $b\left(  r\right)
=2MG$ and $r\in\left[  r_{t},+\infty\right)  $, while for a naked singularity
we have $b\left(  r\right)  =-2\bar{M}G$ and $r\in\left(  0,+\infty\right)  $.
Of course, we are taking under consideration the simplest naked singularity.
For example, one could also consider a Reissner-Nordstr\"{o}m naked
singularity or a Kerr naked singularity. However, the introduction of the
charge in the former and rotation in the latter increase the technical level
and momentarily they will not be considered. On the other hand, when a
Noncommutative geometry is taken under consideration, the spacetime is endowed
with a commutator $\left[  \mathbf{x}^{\mu},\mathbf{x}^{\nu}\right]
=i\,\theta^{\mu\nu}$, where $\theta^{\mu\nu}$ is an antisymmetric matrix which
determines the fundamental discretization of spacetime. As shown in
Ref.\cite{RGPN} and references therein, the classical Liouville measure%
\begin{equation}
\frac{d^{3}\vec{x}d^{3}\vec{k}}{\left(  2\pi\right)  ^{3}}%
\end{equation}
is distorted into%
\begin{equation}
\frac{d^{3}\vec{x}d^{3}\vec{k}}{\left(  2\pi\right)  ^{3}}\exp\left(
-\frac{\theta}{4}k_{i}^{2}\right)  ,\label{NCdn}%
\end{equation}
where $k_{i}^{2}$ is the radial wave number associated to each mode of the
graviton. It is clear that the UV cut off $\theta$ is triggered only by higher
momenta modes $\gtrsim1/\sqrt{\theta}$ which propagate over the background
geometry. In a series of papers\cite{RemoPLB,GaMa,Remof(R)}, we have applied
the Gravity's Rainbow formalism to the Zero Point Energy (ZPE) calculation and
we have shown that appropriate choices of the Rainbow's functions keep under
control the UV divergences. The same finite result has been obtained in
Ref.\cite{RGPN} with a Noncommutative geometry. The key point is the following
expectation value\cite{Remo}\footnote{An application of this calculation in
the framework of Hor\v{a}va-Lifshitzs theory can be found in Ref.\cite{RemoHL}%
. A slight variant of this calculation can be found in Ref. \cite{RemoCharge}%
.}%
\begin{equation}
\frac{1}{V}\frac{\left\langle \Psi\left\vert \int_{\Sigma}d^{3}x\hat{\Lambda
}_{\Sigma}\right\vert \Psi\right\rangle }{\left\langle \Psi|\Psi\right\rangle
}=-\frac{\Lambda}{8\pi G},\label{VEVO}%
\end{equation}
which is obtained by a formal manipulation of the Wheeler-DeWitt equation
(WDW)\cite{DeWitt}. $\Lambda$ denotes the cosmological constant, while
$\hat{\Lambda}_{\Sigma}$ is the operator containing all the information about
the gravitational field. In this form, Eq.$\left(  \ref{VEVO}\right)  $ can be
used to compute ZPE provided that $\Lambda/8\pi G$ be considered as an
eigenvalue of $\hat{\Lambda}_{\Sigma}$. Nevertheless, solving Eq.$\left(
\ref{VEVO}\right)  $ is a quite impossible task, therefore we are oriented to
use a variational approach with trial wave functionals. The related boundary
conditions are dictated by the choice of the trial wave functionals which, in
our case\textbf{,} are of the Gaussian type: this choice is justified by the
fact that ZPE should be described by a good candidate of the \textquotedblleft%
\textit{vacuum state}\textquotedblright. However if we change the form of the
wave functionals we also change the corresponding boundary conditions and
therefore the description of the vacuum state. It is better to observe that
the obtained eigenvalue $\Lambda/8\pi G$, is far to be a constant, rather it
will be dependent on some parameters which depend on the background under
consideration. Therefore the correct interpretation is that of a
\textquotedblleft\textit{dynamical cosmological constant}\textquotedblright%
\ evolving in $r$ and $M$ instead of a temporal parameter $t$. This is not a
novelty, almost all the inflationary models try to substitute a cosmological
constant $\Lambda$ with some fields that change with time. In this case, it is
the gravity itself that gives a dynamical aspect to the \textquotedblleft%
\textit{cosmological constant} $\Lambda$\textquotedblright, or more correctly
the ZPE $\Lambda/8\pi G$, without introducing any kind of external field but
only quantum fluctuations of the \textit{pure gravitational field}. Note that
in this approach it will be the \textquotedblleft\textit{dynamical
cosmological constant}\textquotedblright\ that will give information about the
naked singularity. It is important to remark that we will not follow a
collapsing star or a shell into a naked singularity, but we will consider a
naked singularity which is already existing and motivated by the results
obtained in Refs.\cite{RemoPLB,GaMa,Remof(R),RGPN}, we would like to extend
the same ZPE calculation to a naked singularity of the form%
\begin{equation}
ds^{2}=-N^{2}\left(  r\right)  dt^{2}+\frac{dr^{2}}{1+\frac{2\bar{M}G}{r}%
}+r^{2}\left(  d\theta^{2}+\sin^{2}\theta d\phi^{2}\right)  ,\label{Line}%
\end{equation}
which, in the case of Gravity's Rainbow will be distorted into the line
element $\left(  \ref{LineGRw}\right)  $, while for the Noncommutative
geometry will remain same as described by Eq.$\left(  \ref{Line}\right)  $.
The starting point of our analysis will be the line element $\left(
\ref{Line}\right)  $, which will also be our cornerstone of the whole paper
which is organized as follows. In section \ref{p1}, we derive Eq.$\left(
\ref{VEVO}\right)  $ and we extract the graviton one loop contribution to ZPE
with respect to the desired background. In section \ref{p2}, we report the
results of Ref.\cite{GaMa} with the help of the background $\left(
\ref{Line}\right)  $ adapted for the naked singularity in a Gravity's Rainbow
environment. In section \ref{p3}, we report the results of Ref.\cite{RGPN}
with the help of the background $\left(  \ref{Line}\right)  $ adapted for the
naked singularity in a Noncommutative environment. We summarize and conclude
in section \ref{p4}. Units in which $\hbar=c=k=1$ are used throughout the paper.

\section{Setting Up the ZPE Calculation from the WDW Equation}

\label{p1}In this section we derive the general form for the ZPE calculation
on a spherical symmetric background. The procedure relies heavily on the
formalism outlined in Refs.\cite{RemoPoS,GaMa,RGPN}. The key point for the
derivation is in the Arnowitt-Deser-Misner ($\mathcal{ADM}$)
decomposition\cite{ADM} of space time based on the following line element%
\begin{equation}
ds^{2}=g_{\mu\nu}\left(  x\right)  dx^{\mu}dx^{\nu}=\left(  -N^{2}+N_{i}%
N^{i}\right)  dt^{2}+2N_{j}dtdx^{j}+g_{ij}dx^{i}dx^{j},
\end{equation}
where $N$ is the \textit{lapse }function and $N_{i}$ the \textit{shift
}function. In terms of the $\mathcal{ADM}$ variables, the four dimensional
scalar curvature $\mathcal{R}$ can be decomposed in the following way%
\begin{equation}
\mathcal{R}=R+K_{ij}K^{ij}-\left(  K\right)  ^{2}-2\nabla_{\mu}\left(
Ku^{\mu}+a^{\mu}\right)  , \label{R}%
\end{equation}
where%
\begin{equation}
K_{ij}=-\frac{1}{2N}\left[  \partial_{t}g_{ij}-N_{i|j}-N_{j|i}\right]
\end{equation}
is the second fundamental form, $K=$ $g^{ij}K_{ij}$ is its trace, $R$ is the
three dimensional scalar curvature and $\sqrt{g}$ is the three dimensional
determinant of the metric. The last term in $\left(  \ref{R}\right)  $
represents the boundary terms contribution where the four-velocity $u^{\mu}$
is the timelike unit vector normal to the spacelike hypersurfaces (t=constant)
denoted by $\Sigma_{t}$ and $a^{\mu}=u^{\alpha}\nabla_{\alpha}u^{\mu}$ is the
acceleration of the timelike normal $u^{\mu}$. Thus%
\begin{equation}
\mathcal{L}\left[  N,N_{i},g_{ij}\right]  =\sqrt{-\text{\/\thinspace
\thinspace}^{4}\text{\/{}\negthinspace}g}\left(  \mathcal{R}-2\Lambda\right)
=\frac{N}{2\kappa}\sqrt{g}\text{ }\left[  K_{ij}K^{ij}-K^{2}+\,R-2\Lambda
-2\nabla_{\mu}\left(  Ku^{\mu}+a^{\mu}\right)  \right]  \label{Lag}%
\end{equation}
represents the gravitational Lagrangian density where $\kappa=8\pi G$. After a
Legendre transformation, the WDW equation simply becomes%
\begin{equation}
\mathcal{H}\Psi=\left[  \left(  2\kappa\right)  G_{ijkl}\pi^{ij}\pi^{kl}%
-\frac{\sqrt{g}}{2\kappa}\!{}\!\left(  \,\!R-2\Lambda\right)  \right]  \Psi=0,
\label{WDWO}%
\end{equation}
where $G_{ijkl}$ is the super-metric and where the conjugate super-momentum
$\pi^{ij}$ is defined as%
\begin{equation}
\pi^{ij}=\frac{\delta\mathcal{L}}{\delta\left(  \partial_{t}g_{ij}\right)
}=\left(  g^{ij}K-K^{ij}\text{ }\right)  \frac{\sqrt{g}}{2\kappa}. \label{mom}%
\end{equation}
Note that $\mathcal{H}=0$ represents the classical constraint which guarantees
the invariance under time reparametrization. The other classical constraint
represents the invariance by spatial diffeomorphism and it is described by
$\pi_{|j}^{ij}=0$, where the vertical stroke \textquotedblleft%
$\vert$%
\textquotedblright\ denotes the covariant derivative with respect to the $3D$
metric $g_{ij}$. To reproduce Eq.$\left(  \ref{VEVO}\right)  $ we have to
multiply Eq.$\left(  \ref{WDWO}\right)  $ by $\Psi^{\ast}\left[
g_{ij}\right]  $ and functionally integrate over the three spatial metric
$g_{ij}$. Then by defining the volume of the hypersurface $\Sigma$ as%
\begin{equation}
V=\int_{\Sigma}d^{3}x\sqrt{g}%
\end{equation}
and%
\begin{equation}
\hat{\Lambda}_{\Sigma}=\left(  2\kappa\right)  G_{ijkl}\pi^{ij}\pi^{kl}%
-\sqrt{g}R/\left(  2\kappa\right)  ,
\end{equation}
we arrive at
\begin{equation}
\frac{1}{V}\frac{\int\mathcal{D}\left[  g_{ij}\right]  \Psi^{\ast}\left[
g_{ij}\right]  \int_{\Sigma}d^{3}x\hat{\Lambda}_{\Sigma}\Psi\left[
g_{ij}\right]  }{\int\mathcal{D}\left[  g_{ij}\right]  \Psi^{\ast}\left[
g_{ij}\right]  \Psi\left[  g_{ij}\right]  }=-\frac{\Lambda}{8\pi G},
\label{VEV}%
\end{equation}
namely Eq.$\left(  \ref{VEVO}\right)  $. To further proceed, we consider%
\begin{equation}
g_{ij}=\bar{g}_{ij}+h_{ij},
\end{equation}
where $\bar{g}_{ij}$ is the background metric and $h_{ij}$ is a quantum
fluctuation around a background. However, to extract the graviton
contribution, we also need an orthogonal decomposition on the tangent space of
3-metric deformations \cite{Vassilevich}
\begin{equation}
h_{ij}=\frac{1}{3}\left(  \sigma+2\nabla\cdot\xi\right)  g_{ij}+\left(
L\xi\right)  _{ij}+h_{ij}^{\bot}. \label{p21a}%
\end{equation}
The operator $L$ maps the gauge vector $\xi_{i}$ into symmetric tracefree
tensors
\begin{equation}
\left(  L\xi\right)  _{ij}=\nabla_{i}\xi_{j}+\nabla_{j}\xi_{i}-\frac{2}%
{3}g_{ij}\left(  \nabla\cdot\xi\right)  ,
\end{equation}
$h_{ij}^{\bot}$ is the traceless-transverse component of the perturbation
(TT), namely
\begin{equation}
g^{ij}h_{ij}^{\bot}=0,\qquad\nabla^{i}h_{ij}^{\bot}=0
\end{equation}
and $h$ is the trace of $h_{ij}$. It is immediate to recognize that the trace
element $\sigma=h-2\left(  \nabla\cdot\xi\right)  $ is gauge invariant. If we
perform the same decomposition also on the momentum $\pi^{ij}$, up to second
order Eq.$\left(  \ref{VEV}\right)  $ becomes
\begin{equation}
\frac{1}{V}\frac{\left\langle \Psi\left\vert \int_{\Sigma}d^{3}x\left[
\hat{\Lambda}_{\Sigma}^{\bot}+\hat{\Lambda}_{\Sigma}^{\xi}+\hat{\Lambda
}_{\Sigma}^{\sigma}\right]  ^{\left(  2\right)  }\right\vert \Psi\right\rangle
}{\left\langle \Psi|\Psi\right\rangle }=-\frac{\Lambda}{\kappa}.
\label{lambda0_2}%
\end{equation}
Concerning the measure appearing in $\left(  \ref{VEV}\right)  $, we have to
note that the decomposition $\left(  \ref{p21a}\right)  $ induces the
following transformation on the functional measure $\mathcal{D}h_{ij}%
\rightarrow\mathcal{D}h_{ij}^{\bot}\mathcal{D}\xi_{i}\mathcal{D}\sigma J_{1}$,
where the Jacobian related to the gauge vector variable $\xi_{i}$ is%
\begin{equation}
J=\left[  \det\left(  \bigtriangleup g^{ij}+\frac{1}{3}\nabla^{i}\nabla
^{j}-R^{ij}\right)  \right]  ^{\frac{1}{2}}.
\end{equation}
This is nothing but the famous Faddeev-Popov determinant. It becomes more
transparent if $\xi_{a}$ is further decomposed into a transverse part $\xi
_{a}^{T}$ with $\nabla^{a}\xi_{a}^{T}=0$ and a longitudinal part $\xi
_{a}^{\parallel}$ with $\xi_{a}^{\parallel}=$ $\nabla_{a}\psi$. Then $J$ can
be expressed by an upper triangular matrix for certain backgrounds (e.g.
Schwarzschild in three dimensions). It is immediate to recognize that for an
Einstein space in any dimension, cross terms vanish and $J$ can be expressed
by a block diagonal matrix. Since $\det AB=\det A\det B$, the functional
measure $\mathcal{D}h_{ij}$ factorizes into%
\begin{equation}
\mathcal{D}h_{ij}=\left(  \det\bigtriangleup_{V}^{T}\right)  ^{\frac{1}{2}%
}\left(  \det\left[  \frac{2}{3}\bigtriangleup^{2}+\nabla_{i}R^{ij}\nabla
_{j}\right]  \right)  ^{\frac{1}{2}}\mathcal{D}h_{ij}^{\bot}\ \mathcal{D}%
\xi^{T}\ \mathcal{D}\psi
\end{equation}
leading to the Faddeev-Popov determinant with $\left(  \bigtriangleup_{V}%
^{ij}\right)  ^{T}=\bigtriangleup g^{ij}-R^{ij}$ acting on transverse vectors.
Thus the inner product can be written as%
\begin{align}
&  \int\mathcal{D}h_{ij}^{\bot}\mathcal{D}\xi^{T}\mathcal{D}\sigma\Psi^{\ast
}\left[  h_{ij}^{\bot}\right]  \Psi^{\ast}\left[  \xi^{T}\right]  \Psi^{\ast
}\left[  \sigma\right]  \Psi\left[  h_{ij}^{\bot}\right]  \Psi\left[  \xi
^{T}\right] \nonumber\\
&  \times\ \Psi\left[  \sigma\right]  \left(  \det\bigtriangleup_{V}%
^{T}\right)  ^{\frac{1}{2}}\left(  \det\left[  \frac{2}{3}\bigtriangleup
^{2}+\nabla_{i}R^{ij}\nabla_{j}\right]  \right)  ^{\frac{1}{2}}.
\end{align}
Nevertheless, since there is no interaction between ghost fields and the other
components of the perturbation at this level of approximation, the Jacobian
appearing in the numerator and in the denominator simplify. The reason can be
found in terms of connected and disconnected terms. The disconnected terms
appear in the Faddeev-Popov determinant and above ones are not linked by the
Gaussian integration. This means that disconnected terms in the numerator and
the same ones appearing in the denominator cancel out. Therefore, $\left(
\ref{lambda0_2}\right)  $ factorizes into three pieces. The piece containing
$E_{\Sigma}^{\bot}$, the contribution of the transverse-traceless tensors
(TT), is essentially the graviton contribution representing true physical
degrees of freedom. Regarding the vector operator $\hat{\Lambda}_{\Sigma}^{T}%
$, we observe that under the action of infinitesimal diffeomorphism generated
by a vector field $\epsilon_{i}$, the components of $\left(  \ref{p21a}%
\right)  $ transform as follows \cite{Vassilevich}%
\begin{equation}
\xi_{j}\longrightarrow\xi_{j}+\epsilon_{j},\qquad h\longrightarrow
h+2\nabla\cdot\epsilon,\qquad h_{ij}^{\bot}\longrightarrow h_{ij}^{\bot}.
\label{Gauge}%
\end{equation}
The Killing vectors satisfying the condition $\nabla_{i}\epsilon_{j}%
+\nabla_{j}\epsilon_{i}=0,$ do not change $h_{ij}$, and thus should be
excluded from the gauge group. All other diffeomorphisms act on $h_{ij}$
nontrivially. We need to fix the residual gauge freedom on the vector $\xi
_{i}$. The simplest choice is $\xi_{i}=0.$ This new gauge fixing produces the
same Faddeev-Popov determinant connected to the Jacobian $J$ and therefore
will not contribute to the final value. We are left with%
\begin{equation}
\frac{1}{V}\frac{\left\langle \Psi^{\bot}\left\vert \int_{\Sigma}d^{3}x\left[
\hat{\Lambda}_{\Sigma}^{\bot}\right]  ^{\left(  2\right)  }\right\vert
\Psi^{\bot}\right\rangle }{\left\langle \Psi^{\bot}|\Psi^{\bot}\right\rangle
}+\frac{1}{V}\frac{\left\langle \Psi^{\sigma}\left\vert \int_{\Sigma}%
d^{3}x\left[  \hat{\Lambda}_{\Sigma}^{\sigma}\right]  ^{\left(  2\right)
}\right\vert \Psi^{\sigma}\right\rangle }{\left\langle \Psi^{\sigma}%
|\Psi^{\sigma}\right\rangle }=-\frac{\Lambda}{\kappa}. \label{lambda0_2a}%
\end{equation}
Note that in the expansion of $\int_{\Sigma}d^{3}x\sqrt{g}{}R$ to second
order, a coupling term between the TT component and scalar one remains. The
scalar contribution $\hat{\Lambda}_{\Sigma}^{\sigma}$ can be always gauged
away by an appropriate choice of the vector field $\epsilon_{j}$. Now that we
have deduced the one loop approximation of Eq.$\left(  \ref{VEVO}\right)  $,
we need a regularization/renormalization process to keep under control the
divergences. In the next section we will evaluate Eq.$\left(  \ref{lambda0_2a}%
\right)  $ distorted by Gravity's Rainbow which will be our regularization framework.

\section{Setting up the ZPE Computation with the Wheeler-DeWitt Equation
distorted by Gravity's Rainbow}

\label{p2}In this section we derive how WDW modifies when the functions
$g_{1}\left(  E/E_{P}\right)  $ and $g_{2}\left(  E/E_{P}\right)  $ distort
the background $\left(  \ref{Line}\right)  $. The form of the background is
such that the \textit{shift function}%
\begin{equation}
N^{i}=-Nu^{i}=g_{0}^{4i}=0
\end{equation}
vanishes, while $N$ is the previously defined \textit{lapse function}. Thus
the definition of $K_{ij}$ implies%
\begin{equation}
K_{ij}=-\frac{\dot{g}_{ij}}{2N}=\frac{g_{1}\left(  E/E_{P}\right)  }{g_{2}%
^{2}\left(  E/E_{P}\right)  }\tilde{K}_{ij},\label{Kij}%
\end{equation}
where the dot denotes differentiation with respect to the time $t$ and the
tilde indicates the quantity computed in absence of rainbow's functions
$g_{1}\left(  E/E_{P}\right)  $ and $g_{2}\left(  E/E_{P}\right)  $. The trace
of the extrinsic curvature, therefore becomes%
\begin{equation}
K=g^{ij}K_{ij}=g_{1}\left(  E/E_{P}\right)  \tilde{K}%
\end{equation}
and the momentum $\pi^{ij}$ conjugate to the three-metric $g_{ij}$ of $\Sigma$
is%
\begin{equation}
\pi^{ij}=\frac{\sqrt{g}}{2\kappa}\left(  Kg^{ij}-K^{ij}\right)  =\frac
{g_{1}\left(  E/E_{P}\right)  }{g_{2}\left(  E/E_{P}\right)  }\tilde{\pi}%
^{ij}.
\end{equation}
Thus the distorted classical constraint becomes%
\begin{equation}
\mathcal{H}=\left(  2\kappa\right)  \frac{g_{1}^{2}\left(  E/E_{P}\right)
}{g_{2}^{3}\left(  E/E_{P}\right)  }\tilde{G}_{ijkl}\tilde{\pi}^{ij}\tilde
{\pi}^{kl}\mathcal{-}\frac{\sqrt{\tilde{g}}}{2\kappa g_{2}\left(
E/E_{P}\right)  }\!{}\!\left(  \tilde{R}-\frac{2\Lambda_{c}}{g_{2}^{2}\left(
E/E_{P}\right)  }\right)  =0,\label{Acca}%
\end{equation}
where we have used the following property on $R$%
\begin{equation}
R=g^{ij}R_{ij}=g_{2}^{2}\left(  E/E_{P}\right)  \tilde{R}%
\end{equation}
and where
\begin{equation}
G_{ijkl}=\frac{1}{2\sqrt{g}}\left(  g_{ik}g_{jl}+g_{il}g_{jk}-g_{ij}%
g_{kl}\right)  =\frac{\tilde{G}_{ijkl}}{g_{2}\left(  E/E_{P}\right)  }.
\end{equation}
The corresponding vacuum expectation value $\left(  \ref{VEVO}\right)  $
becomes%
\begin{equation}
\frac{g_{2}^{3}\left(  E/E_{P}\right)  }{\tilde{V}}\frac{\left\langle
\Psi\left\vert \int_{\Sigma}d^{3}x\tilde{\Lambda}_{\Sigma}\right\vert
\Psi\right\rangle }{\left\langle \Psi|\Psi\right\rangle }=-\frac{\Lambda
}{\kappa},\label{WDW1}%
\end{equation}
with%
\begin{equation}
\tilde{\Lambda}_{\Sigma}=\left(  2\kappa\right)  \frac{g_{1}^{2}\left(
E/E_{P}\right)  }{g_{2}^{3}\left(  E/E_{P}\right)  }\tilde{G}_{ijkl}\tilde
{\pi}^{ij}\tilde{\pi}^{kl}\mathcal{-}\frac{\sqrt{\tilde{g}}\tilde{R}}{\left(
2\kappa\right)  g_{2}\left(  E/E_{P}\right)  }\!{}\!.\label{LambdaR}%
\end{equation}
Extracting the TT tensor contribution from Eq.$\left(  \ref{WDW1}\right)  $,
we find%
\begin{equation}
\hat{\Lambda}_{\Sigma}^{\bot}=\frac{g_{2}^{3}\left(  E/E_{P}\right)  }%
{4\tilde{V}}\int_{\Sigma}d^{3}x\sqrt{\overset{\sim}{\bar{g}}}\tilde{G}%
^{ijkl}\left[  \left(  2\kappa\right)  \frac{g_{1}^{2}\left(  E/E_{P}\right)
}{g_{2}^{3}\left(  E/E_{P}\right)  }\tilde{K}^{-1\bot}\left(  x,x\right)
_{ijkl}+\frac{1}{\left(  2\kappa\right)  g_{2}\left(  E/E_{P}\right)  }%
\!{}\left(  \tilde{\bigtriangleup}_{L\!}^{m}\tilde{K}^{\bot}\left(
x,x\right)  \right)  _{ijkl}\right]  ,\label{p22}%
\end{equation}
with the prescription that the corresponding eigenvalue equation transforms
into the following way%
\begin{equation}
\left(  \hat{\bigtriangleup}_{L\!}^{m}\!{}h^{\bot}\right)  _{ij}=E^{2}%
h_{ij}^{\bot}\qquad\rightarrow\qquad\left(  \tilde{\bigtriangleup}_{L\!}%
^{m}\!{}\tilde{h}^{\bot}\right)  _{ij}\!{}=\frac{E^{2}}{g_{2}^{2}\left(
E/E_{P}\right)  }\tilde{h}_{ij}^{\bot}\label{EE}%
\end{equation}
in order to reestablish the correct way of transformation of the perturbation.
Eq.$\left(  \ref{EE}\right)  $ is the equation connecting the graviton energy
with Gravity's Rainbow.\textbf{ }The propagator $K^{\bot}\left(  x,x\right)
_{iakl}$ will transform as
\begin{equation}
K^{\bot}\left(  \overrightarrow{x},\overrightarrow{y}\right)  _{iakl}%
\rightarrow\frac{1}{g_{2}^{4}\left(  E/E_{P}\right)  }\tilde{K}^{\bot}\left(
\overrightarrow{x},\overrightarrow{y}\right)  _{iakl}.\label{proptt}%
\end{equation}
Thus the total one loop energy density for the graviton for the distorted GR
becomes%
\begin{equation}
\frac{\Lambda}{8\pi G}=-\frac{1}{2\tilde{V}}\sum_{\tau}g_{1}\left(
E/E_{P}\right)  g_{2}\left(  E/E_{P}\right)  \left[  \sqrt{E_{1}^{2}\left(
\tau\right)  }+\sqrt{E_{2}^{2}\left(  \tau\right)  }\right]  .\label{VEVR}%
\end{equation}
The above expression makes sense only for $E_{i}^{2}\left(  \tau\right)  >0$,
where $E_{i}$ are the eigenvalues of $\tilde{\bigtriangleup}_{L\!}^{m}$. With
the help of Regge and Wheeler representation\cite{Regge Wheeler}, the
eigenvalue equation $\left(  \ref{EE}\right)  $ can be reduced to%
\begin{equation}
\left[  -\frac{d^{2}}{dx^{2}}+\frac{l\left(  l+1\right)  }{r^{2}}+m_{i}%
^{2}\left(  r\right)  \right]  f_{i}\left(  x\right)  =\frac{E_{i,l}^{2}%
}{g_{2}^{2}\left(  E/E_{P}\right)  }f_{i}\left(  x\right)  \quad
i=1,2\quad,\label{p34}%
\end{equation}
where we have used reduced fields of the form $f_{i}\left(  x\right)
=F_{i}\left(  x\right)  /r$ and where we have defined two r-dependent
effective masses $m_{1}^{2}\left(  r\right)  $ and $m_{2}^{2}\left(  r\right)
$%
\begin{equation}
\left\{
\begin{array}
[c]{c}%
m_{1}^{2}\left(  r\right)  =\frac{6}{r^{2}}\left(  1-\frac{b\left(  r\right)
}{r}\right)  +\frac{3}{2r^{2}}b^{\prime}\left(  r\right)  -\frac{3}{2r^{3}%
}b\left(  r\right)  \\
\\
m_{2}^{2}\left(  r\right)  =\frac{6}{r^{2}}\left(  1-\frac{b\left(  r\right)
}{r}\right)  +\frac{1}{2r^{2}}b^{\prime}\left(  r\right)  +\frac{3}{2r^{3}%
}b\left(  r\right)
\end{array}
\right.  \quad\left(  r\equiv r\left(  x\right)  \right)  .\label{masses}%
\end{equation}
In order to use the W.K.B. approximation, from Eq.$\left(  \ref{p34}\right)  $
we can extract two r-dependent radial wave numbers%
\begin{equation}
k_{i}^{2}\left(  r,l,\omega_{i,nl}\right)  =\frac{E_{i,nl}^{2}}{g_{2}%
^{2}\left(  E/E_{P}\right)  }-\frac{l\left(  l+1\right)  }{r^{2}}-m_{i}%
^{2}\left(  r\right)  \quad i=1,2\quad.
\end{equation}
To further proceed we use the W.K.B. method used by `t Hooft in the brick wall
problem\cite{tHooft} and we count the number of modes with frequency less than
$\omega_{i}$, $i=1,2$. This is given approximately by%
\begin{equation}
\tilde{g}\left(  E_{i}\right)  =\int_{0}^{l_{\max}}\nu_{i}\left(
l,E_{i}\right)  \left(  2l+1\right)  dl,\label{p41}%
\end{equation}
where $\nu_{i}\left(  l,E_{i}\right)  $, $i=1,2$ is the number of nodes in the
mode with $\left(  l,E_{i}\right)  $, such that $\left(  r\equiv r\left(
x\right)  \right)  $
\begin{equation}
\nu_{i}\left(  l,E_{i}\right)  =\frac{1}{\pi}\int_{-\infty}^{+\infty}%
dx\sqrt{k_{i}^{2}\left(  r,l,E_{i}\right)  }.\label{p42}%
\end{equation}
Here it is understood that the integration with respect to $x$ and $l_{\max}$
is taken over those values which satisfy $k_{i}^{2}\left(  r,l,E_{i}\right)
\geq0,$ $i=1,2$. With the help of Eqs.$\left(  \ref{p41},\ref{p42}\right)  $,
Eq.$\left(  \ref{VEVR}\right)  $ leads to%
\begin{equation}
\frac{\Lambda}{8\pi G}=-\frac{1}{\pi}\sum_{i=1}^{2}\int_{0}^{+\infty}%
E_{i}g_{1}\left(  E/E_{P}\right)  g_{2}\left(  E/E_{P}\right)  \frac
{d\tilde{g}\left(  E_{i}\right)  }{dE_{i}}dE_{i}.\label{tot1loop}%
\end{equation}
This is the graviton contribution to the induced cosmological constant to one
loop. The explicit evaluation of the density of states yields%
\[
\frac{d\tilde{g}(E_{i})}{dE_{i}}=\int\frac{\partial\nu(l{,}E_{i})}{\partial
E_{i}}(2l+1)dl=\frac{1}{\pi}\int_{-\infty}^{+\infty}dx\int_{0}^{l_{\max}}%
\frac{(2l+1)}{\sqrt{k^{2}(r,l,E)}}\frac{d}{dE_{i}}\left(  \frac{E_{i}^{2}%
}{g_{2}^{2}\left(  E/E_{P}\right)  }-m_{i}^{2}\left(  r\right)  \right)  dl
\]%
\begin{equation}
=\frac{4}{3\pi}\int_{-\infty}^{+\infty}dxr^{2}\frac{d}{dE_{i}}\left(
\frac{E_{i}^{2}}{g_{2}^{2}\left(  E/E_{P}\right)  }-m_{i}^{2}\left(  r\right)
\right)  ^{\frac{3}{2}}.\label{states}%
\end{equation}
Plugging expression $\left(  \ref{states}\right)  $ into Eq.$\left(
\ref{tot1loop}\right)  $ and dividing for a volume factor, we obtain%
\begin{equation}
\frac{\Lambda}{8\pi G}=-\frac{1}{3\pi^{2}}\sum_{i=1}^{2}\int_{E^{\ast}%
}^{+\infty}E_{i}g_{1}\left(  E/E_{P}\right)  g_{2}\left(  E/E_{P}\right)
\frac{d}{dE_{i}}\sqrt{\left(  \frac{E_{i}^{2}}{g_{2}^{2}\left(  E/E_{P}%
\right)  }-m_{i}^{2}\left(  r\right)  \right)  ^{3}}dE_{i},\label{LoverG}%
\end{equation}
where $E^{\ast}$ is the value which annihilates the argument of the root. In
the previous equation, we have included an additional $4\pi$ factor coming
from the angular integration and we have assumed that the effective mass does
not depend on the energy $E$. It is immediate to recognize that not every form
of $g_{1}\left(  E/E_{P}\right)  $ and $g_{2}\left(  E/E_{P}\right)  $ can be
used to compute the integrals in Eq.$\left(  \ref{LoverG}\right)  $. Indeed,
we need to impose that the Rainbow's functions satisfy convergence criteria.
We fix our attention on the following choice
\begin{equation}
g_{1}(E/E_{P})=\left(  1+\beta\frac{E}{E_{P}}\right)  \exp\left(  -\alpha
\frac{E^{2}}{E_{P}^{2}}\right)  ~~~~~~~~\text{and}~~~~~~~~g_{2}(E/E_{P}%
)=1,\label{GRw}%
\end{equation}
which has been extensively used in Refs.\cite{GaMa}. For the Schwarzschild
case, the background satisfies the following property%
\begin{equation}
m_{0}^{2}\left(  r\right)  =m_{2}^{2}\left(  r\right)  =-m_{1}^{2}\left(
r\right)  ,\qquad\forall r\in\left(  r_{t},r_{1}\right)  ,\label{cond}%
\end{equation}
with $r_{t}=2MG$ and $r\in\left[  r_{t},5r_{t}/2\right]  $ and for the dS, AdS
and Minkowski background \textbf{,} the property%
\begin{equation}
m_{0}^{2}\left(  r\right)  =m_{2}^{2}\left(  r\right)  =m_{1}^{2}\left(
r\right)  ,\qquad\forall r\in\left(  0,\infty\right)  \label{equal}%
\end{equation}
is satisfied. So in the case of naked singularity, we find%
\begin{equation}
m_{1}^{2}(r)=\frac{6}{r^{2}}+\frac{15\bar{M}G}{r^{3}}\label{m1}%
\end{equation}
and
\begin{equation}
m_{2}^{2}(r)=\frac{6}{r^{2}}+\frac{9\bar{M}G}{r^{3}},\label{m2}%
\end{equation}
with $\bar{M}=-M$. Eq.$\left(  \ref{LoverG}\right)  $ becomes%
\begin{equation}
\frac{\Lambda}{8\pi G}=-\frac{1}{3\pi^{2}}\left(  I_{1}+I_{2}\right)
,\label{LoveG}%
\end{equation}
where%
\begin{equation}
I_{1,2}=\int_{\sqrt{m_{1,2}^{2}\left(  r\right)  }}^{\infty}E^{2}\left(
1+\beta\frac{E}{E_{P}}\right)  \exp\left(  -\alpha\frac{E^{2}}{E_{P}^{2}%
}\right)  \sqrt{E^{2}-m_{1,2}^{2}\left(  r\right)  }dE.\label{Ia}%
\end{equation}
Using the results of appendix \ref{app1}, the integrals can be evaluated and
we can finally write%
\[
\frac{\Lambda}{8\pi G}=-\frac{E_{P}^{4}}{8\pi^{2}\alpha^{2}}\left[
\exp\left(  -x^{2}\alpha\right)  \beta\sqrt{\pi}\frac{3+2x^{2}\alpha}%
{\sqrt{\alpha}}+2\alpha x^{2}\exp\left(  -\frac{x^{2}\alpha}{2}\right)
K_{1}\left(  \frac{x^{2}\alpha}{2}\right)  \right.
\]%
\begin{equation}
\left.  +\exp\left(  -y^{2}\alpha\right)  \beta\sqrt{\pi}\frac{3+2y^{2}\alpha
}{\sqrt{\alpha}}+2\alpha y^{2}\exp\left(  -\frac{y^{2}\alpha}{2}\right)
K_{1}\left(  \frac{y^{2}\alpha}{2}\right)  \right]  ,\label{LG}%
\end{equation}
where%
\begin{equation}
x=\sqrt{\frac{m_{1}^{2}(r)}{E_{P}^{2}}}=\frac{1}{rE_{P}}\sqrt{6+\frac
{15\bar{M}G}{r}}\qquad\mathrm{and\qquad}y=\sqrt{\frac{m_{2}^{2}(r)}{E_{P}^{2}%
}}=\frac{1}{rE_{P}}\sqrt{6+\frac{9\bar{M}G}{r}}.\label{xy}%
\end{equation}
It is useful to see what happens when $x$ and $y\rightarrow\infty$ in
Eq.$\left(  \ref{LG}\right)  $. Taking the leading term, one gets%
\begin{equation}
\frac{\Lambda}{8\pi GE_{P}^{4}}=-\frac{\beta}{2\pi^{3/2}\alpha^{3/2}}\left[
e^{-x^{2}\alpha}x^{2}+e^{-y^{2}\alpha}y^{2}\right]  ,\label{LGLarge}%
\end{equation}
while when $x$ and $y\rightarrow0$, we find%
\[
\frac{\Lambda}{8\pi GE_{P}^{4}}=-\frac{4+3\sqrt{\pi/\alpha}\beta}{4\pi
^{2}\alpha^{2}}+\frac{2+\sqrt{\pi/\alpha}\beta}{8\pi^{2}\alpha}\left(
x^{2}+y^{2}\right)  -\frac{x^{4}}{32\pi^{2}}\log\left(  \frac{x^{4}\alpha
^{2}e^{1+2\gamma-2\sqrt{\pi/\alpha}\beta}}{16}\right)
\]%
\begin{equation}
-\frac{y^{4}}{32\pi^{2}}\log\left(  \frac{y^{4}\alpha^{2}e^{1+2\gamma
-2\sqrt{\pi/\alpha}\beta}}{16}\right)  ~~,\label{LGSmall}%
\end{equation}
where $\gamma$ is Euler's constant. It is immediate to see that if we set%
\begin{equation}
\beta=-\frac{4}{3}\sqrt{\frac{\alpha}{\pi}},\label{beta}%
\end{equation}
\newline then%
\begin{equation}
\lim_{\substack{x\rightarrow0\\y\rightarrow0}}\frac{\Lambda}{8\pi GE_{P}^{4}%
}=-\frac{4+3\sqrt{\pi/\alpha}~\beta}{4\pi^{2}\alpha^{2}}=0.
\end{equation}
Using the explicit form of the variables $x$ and $y$ of Eq.$\left(
\ref{xy}\right)  $ and plugging the value of the parameter $\beta$ found in
Eq.$\left(  \ref{beta}\right)  $ into the asymptotic expansion $\left(
\ref{LGLarge}\right)  $, one finds that the leading term behaves as%
\begin{equation}
\frac{\Lambda}{8\pi GE_{P}^{4}}=\frac{2}{3\pi^{2}\alpha r^{3}E_{P}^{2}}\left[
15\bar{M}G\exp\left(  -\frac{\alpha15\bar{M}G}{r^{3}E_{P}^{2}}\right)
+9\bar{M}G\exp\left(  -\frac{\alpha9\bar{M}G}{r^{3}E_{P}^{2}}\right)  \right]
,
\end{equation}
where it is meant that either $r\rightarrow0$ or $\bar{M}\rightarrow\infty$.
Nevertheless, the case in which $\bar{M}\rightarrow\infty$ is unphysical
because it represents a singularity which fills the whole universe. On the
other hand the case in which $r\rightarrow0$ represents a naked singularity
which is no more singular. In other words the distortion due to Gravity's
Rainbow can cure also the problem of the singularity which appears appropriate
for a correct theory of Quantum Gravity. Note that the regularity at the
origin has been obtained also for the de Sitter and Anti-de Sitter spaces in
Ref.\cite{GaMa}. It is also important to remark that the regularity at $r=0$,
does not appear for every proposal of the Rainbow's functions $g_{1}\left(
E/E_{P}\right)  $ and $g_{2}\left(  E/E_{P}\right)  $. Indeed the proposal%
\begin{equation}
g_{1}\left(  \frac{E}{E_{P}}\right)  =(1+c_{2}\frac{E}{E_{P}})\exp(-c_{1}%
\frac{E^{2}}{E_{P}^{2}})\qquad g_{2}\left(  E/E_{P}\right)  =1+c_{3}\frac
{E}{E_{P}}\label{b)}%
\end{equation}
discussed in Ref.\cite{Remof(R)} cannot be adopted here because in the
trans-Planckian region the argument of the integrals $\left(  \ref{Ia}\right)
$ become independent on the radial variable and therefore they are not
suppressed near the singularity. Therefore it appears that the choice $\left(
\ref{GRw}\right)  $ is very special in this context. In the next section, we
will compute the effect of a Noncommutative theory on a naked singularity background.

\section{Setting up the ZPE Computation with the Wheeler-DeWitt Equation
distorted by a Noncommutative Geometry}

\label{p3}If we avoid the use of Gravity's Rainbow and we introduce a
Noncommutative Geometry, the first thing to do is the recovery of the one loop
contribution of the graviton%
\begin{equation}
\frac{\Lambda}{8\pi G}=-\frac{1}{2V}\sum_{\tau}\left[  \sqrt{E_{1}^{2}\left(
\tau\right)  }+\sqrt{E_{2}^{2}\left(  \tau\right)  }\right]  . \label{1loop}%
\end{equation}
Eq.$\left(  \ref{1loop}\right)  $ has the same expression of Eq.$\left(
\ref{VEVR}\right)  $, but with $g_{1}\left(  E/E_{P}\right)  =g_{2}\left(
E/E_{P}\right)  =1$. However, to obtain a finite result we need to replace the
classical Liouville counting number of nodes%
\begin{equation}
dn=\frac{d^{3}\vec{x}d^{3}\vec{k}}{\left(  2\pi\right)  ^{3}}%
\end{equation}
with\cite{NCThermo,NCStat}%
\begin{equation}
dn_{i}=\frac{d^{3}\vec{x}d^{3}\vec{k}}{\left(  2\pi\right)  ^{3}}\exp\left(
-\frac{\theta}{4}k_{i}^{2}\right)  , \label{moddn}%
\end{equation}
where%
\begin{equation}
k_{i}^{2}=E_{i,nl}^{2}-m_{i}^{2}\left(  r\right)  \quad i=1,2.
\end{equation}
This deformation corresponds to an effective cut off on the background
geometry $\left(  \ref{Line}\right)  $. The UV cut off is triggered only by
higher momenta modes $\gtrsim1/\sqrt{\theta}$ which propagate over the
background geometry. The virtue of this kind of deformation is its exponential
damping profile, which encodes an intrinsic nonlocal character into fields
$f_{i}(x)$. Plugging $\left(  \ref{p42}\right)  $ into $\left(  \ref{p41}%
\right)  $ and taking account of $\left(  \ref{moddn}\right)  $, the number of
modes with frequency less than $\omega_{i}$, $i=1,2$ is given by%
\begin{align}
&  g\left(  E_{i}\right)  =\frac{1}{\pi}\int_{-\infty}^{+\infty}dx\int
_{0}^{l_{\max}}\sqrt{E_{i,nl}^{2}-\frac{l\left(  l+1\right)  }{r^{2}}%
-m_{i}^{2}\left(  r\right)  }\nonumber\\
&  \times\left(  2l+1\right)  \exp\left(  -\frac{\theta}{4}k_{i}^{2}\right)
\ dl.
\end{align}
After integration over modes, one gets
\begin{align}
&  g\left(  E_{i}\right)  =\frac{2}{3\pi}\int_{-\infty}^{+\infty}%
dx\ r^{2}\left[  \frac{3}{2}\sqrt{\left(  E_{i,nl}^{2}-m_{i}^{2}\left(
r\right)  \right)  ^{3}}\right. \nonumber\\
&  \left.  \exp\left(  -\frac{\theta}{4}\left(  E_{i,nl}^{2}-m_{i}^{2}\left(
r\right)  \right)  \right)  \right]  . \label{g(E)}%
\end{align}
This form of $g\left(  E_{i}\right)  $ allows an integration by parts in
$\left(  \ref{VEVR}\right)  $ leading to%
\begin{equation}
\frac{\Lambda}{8\pi G}=-\frac{1}{4\pi^{2}V}\sum_{i=1}^{2}\int_{0}^{+\infty
}E_{i}\frac{dg\left(  E_{i}\right)  }{dE_{i}}dE_{i}=\frac{1}{4\pi^{2}V}%
\sum_{i=1}^{2}\int_{0}^{+\infty}g\left(  E_{i}\right)  dE_{i}. \label{t1loop}%
\end{equation}
This is the graviton contribution to the induced cosmological constant at one
loop, where an additional $4\pi$ coming from the angular integration has been
included. Plugging Eq.$\left(  \ref{g(E)}\right)  $ into Eq.$\left(
\ref{t1loop}\right)  $, one finds%
\begin{equation}
\frac{\Lambda}{8\pi G}=\frac{1}{6\pi^{2}}\sum_{i=1}^{2}\int_{\sqrt{m_{i}%
^{2}\left(  r\right)  }}^{+\infty}\sqrt{\left(  \omega^{2}-m_{i}^{2}\left(
r\right)  \right)  ^{3}}e^{-\frac{\theta}{4}\left(  \omega^{2}-m_{i}%
^{2}\left(  r\right)  \right)  }, \label{Schw1loop}%
\end{equation}
where $m_{i}^{2}\left(  r\right)  $ are the effective masses described by
Eqs.$\left(  \ref{m1},\ref{m2}\right)  $

Plugging the result of $\left(  \ref{I1}\right)  $ into $\left(
\ref{t1loop}\right)  $, we get%
\begin{align}
\frac{\Lambda}{8\pi G}  &  =\frac{1}{12\pi^{2}}\left(  \frac{4}{\theta
}\right)  ^{2}\left[  \left(  \frac{1}{2}z\left(  1-z\right)  K_{1}\left(
\frac{z}{2}\right)  +\frac{1}{2}z^{2}K_{0}\left(  \frac{z}{2}\right)  \right)
\exp\left(  \frac{z}{2}\right)  \right. \nonumber\\
&  \left.  +\left(  \frac{1}{2}w\left(  1-w\right)  K_{1}\left(  \frac{w}%
{2}\right)  +\frac{1}{2}w^{2}K_{0}\left(  \frac{w}{2}\right)  \right)
\exp\left(  \frac{w}{2}\right)  \right]  \label{LambdaNC}%
\end{align}
where%
\begin{equation}
\left\{
\begin{array}
[c]{c}%
z=m_{1}^{2}\left(  r\right)  \theta/4=\left(  \frac{6}{r^{2}}+\frac{15\bar
{M}G}{r^{3}}\right)  \frac{\theta}{4}\\
w=m_{2}^{2}\left(  r\right)  \theta/4=\left(  \frac{6}{r^{2}}+\frac{9\bar{M}%
G}{r^{3}}\right)  \frac{\theta}{4}%
\end{array}
\right.  .
\end{equation}
To analyze these results we consider the asymptotic expansion for
$z,w\rightarrow\infty$ which means $r\ll\sqrt{\theta}$. Then one gets
\begin{equation}
\frac{\Lambda}{8\pi G}\simeq\frac{1}{12\pi^{2}}\left(  \frac{4}{\theta
}\right)  ^{2}\frac{3}{8}\left(  \sqrt{\frac{\pi}{z}}+\sqrt{\frac{\pi}{w}%
}\right)  \rightarrow0,
\end{equation}
This corresponds to the correct behavior in a spacetime region where the
curvature vanishes. On the other hand, for $r\gg\sqrt{\theta}$ we have
$z,w\rightarrow0$ which implies%
\begin{gather}
\frac{\Lambda}{8\pi G}\simeq\frac{1}{12\pi^{2}}\left(  \frac{4}{\theta
}\right)  ^{2}\left[  2-\frac{z+w}{2}-\frac{3}{8}\ln\left(  \frac
{ze^{\gamma+\frac{7}{6}}}{4}\right)  z^{2}\right. \nonumber\\
-\left.  \frac{3}{8}\ln\left(  \frac{we^{\gamma+\frac{7}{6}}}{4}\right)
w^{2}\right]  \rightarrow\frac{8}{3\pi^{2}\theta^{2}} \label{LNC}%
\end{gather}
i.e. a finite value of the cosmological term.

\section{Conclusions}

\label{p4}In this paper we have considered the ZPE contribution deriving from
an existing naked singularity. As an example we have considered the negative
Schwarzschild mass which is the simplest model of naked singularity. We have
computed the ZPE to one loop which is described as an induced cosmological
constant. To keep under control the UV divergencies, instead of using a
standard regularization/renormalization scheme we have chosen to distort the
gravitational field with the help of two proposals: Gravity's Rainbow and the
Noncommutative geometry. This choice has been motivated by several results
obtained computing the ZPE on some spherically symmetric
background\cite{RemoPLB,GaMa,Remof(R),RGPN}. What we have found is that the
distortion of the gravitational field can eliminate the singularity. In
particular, we find that some Rainbow's functions suppress the divergent
behavior so strongly in such a way to give regularity even to the point $r=0$.
Of course this cannot happen for every choice of the Rainbow's functions, as
shown with the choice $\left(  \ref{b)}\right)  $. It is important to remark
that the choice of $g_{1}\left(  E/E_{P}\right)  $ and $g_{2}\left(
E/E_{P}\right)  $ for a ZPE calculation is restricted by the condition
$\left(  \ref{lim}\right)  $ and by the condition that the integrals for the
graviton to one loop be finite. This means that the choice is not unique.
Indeed in Ref.\cite{RemoPLB}, the adopted choice was%
\begin{equation}
\frac{g_{1}\left(  E/E_{P}\right)  }{g_{2}\left(  E/E_{P}\right)  }%
=\exp\left(  -\frac{E}{E_{P}}\right)  ,\label{rap}%
\end{equation}
\textbf{ }while in this paper, inspired by Noncommutative geometry, we have
chosen%
\begin{equation}
\frac{g_{1}\left(  E/E_{P}\right)  }{g_{2}\left(  E/E_{P}\right)  }=\left(
1+\beta\frac{E}{E_{P}}\right)  \exp\left(  -\alpha\frac{E^{2}}{E_{P}^{2}%
}\right)  .\label{gauss}%
\end{equation}
and both the choices lead to a finite result. Nothing prevents to relax the
condition $\left(  \ref{gauss}\right)  $ into condition $\left(
\ref{rap}\right)  $, but this goes beyond the purpose of this paper. It is
important to remark that in Gravity's Rainbow with the choice $\left(
\ref{GRw}\right)  $, the Minkowski limit test is satisfied. We draw to the
reader's attention that for Minkowski limit we mean the following
prescription\cite{Remof(R)}%
\begin{equation}
\lim_{\bar{M}\rightarrow0}\frac{\Lambda}{8\pi G}=0.\label{MLim}%
\end{equation}
This means that when the background is switched off, one should recover the
features of a Minkowski background. Note that the same test is not passed by a
Noncommutative distortion. Indeed looking at Eq.$\left(  \ref{LNC}\right)  $,
one finds%
\begin{equation}
\lim_{\bar{M}\rightarrow0}\frac{\Lambda}{8\pi G}=\frac{8}{3\pi^{2}\theta^{2}},
\end{equation}
namely the granularity of the Noncommutative geometry persists independently
on the value of the naked singularity. Only when $\theta\rightarrow\infty$, we
have the vanishing limit, but this is a unphysical situation and therefore
will be discarded. One possibility to overcome this difficulty is in a further
modification of the theory coming from the replacement of the $4D$ scalar
curvature $\mathcal{R}$ with an arbitrary function of the scalar curvature,
namely an $f\left(  \mathcal{R}\right)  $ theory.\textbf{ }Actually, one could
introduce complicated combinations including $R^{2}$, $R^{\mu\nu}R_{\mu\nu}$,
$R^{\mu\nu\alpha\beta}R_{\mu\nu\alpha\beta}$, $R\,\Box R$, $R\,\Box^{k}R$.
These modifications are known under the name of Extended Theories of Gravity
(ETG) and they have been introduced to explain data on the Large Scale
Structure of Space-Time\cite{SCMdL}. Since ETG introduce higher curvature
terms, we have a benefit even at short scales where the construction of an
effective action in Quantum Gravity is possible\cite{BOS}. It is interesting
to note that combining the simplest ETG model, namely an $f\left(
\mathcal{R}\right)  $ theory with Gravity's Rainbow, one obtains a model with
interesting features in the Infra-Red and which is finite in the Ultra-Violet
range, at least to one loop\cite{Remof(R)}. Moreover, thanks to the
flexibility of the $f\left(  \mathcal{R}\right)  $ term one can obtain the
appropriate Minkowski limit, in the sense of Eq.$\left(  \ref{MLim}\right)  $.
It is interesting to observe that the same behavior is present for the
Schwarzschild solution. Therefore it seems that the ZPE calculation in the
context of naked singularity with a Gravity's Rainbow distortion appears to be special.

\begin{acknowledgments}
B.M. was partly supported by the Exellence-in-Research Fellowship of IIT Gandhinagar.
\end{acknowledgments}

\appendix{}

\section{Integrals for Gravity's Rainbow distortion}

\label{app1}In this appendix we give the rules to solve the integrals $I_{1}$
and $I_{2}$, given by Eq.$\left(  \ref{Ia}\right)  $ with%
\begin{equation}
g_{1}(\frac{E}{E_{P}})=\left(  1+\beta\frac{E}{E_{P}}\right)  \exp
(-\alpha\frac{E^{2}}{E_{P}^{2}})\,,\qquad g_{2}\left(  \frac{E}{E_{P}}\right)
=1;\qquad\alpha>0,\beta\in\mathbb{R}.
\end{equation}
Changing variables $E=\sqrt{x}$, the first term of the integral in $\left(
\ref{Ia}\right)  $ becomes%
\begin{align}
I  &  =\frac{1}{2}\int_{\sqrt{m^{2}}}^{\infty}\exp(-\alpha\frac{x}{E_{P}^{2}%
})\sqrt{x}\sqrt{x-m^{2}}dx\nonumber\\
&  =\frac{E_{P}^{4}}{2\sqrt{\pi}}\left(  \frac{m^{2}}{\alpha E_{P}^{2}%
}\right)  \Gamma\left(  \frac{3}{2}\right)  \exp\left(  -\frac{\alpha m^{2}%
}{2E_{P}^{2}}\right)  K_{1}\left(  \frac{\alpha m^{2}}{2E_{P}^{2}}\right)  ,
\end{align}
where we have used the following relationship
\begin{equation}
\int_{u}^{\infty}\left(  x-u\right)  ^{\mu-1}x^{\mu-1}\exp\left(  -\beta
x\right)  dx=\frac{\Gamma\left(  \mu\right)  }{\sqrt{\pi}}\left(  \frac
{u}{\beta}\right)  ^{\mu-1/2}\exp\left(  -\frac{\beta u}{2}\right)
K_{\mu-1/2}\left(  \frac{\beta u}{2}\right)  \qquad%
\begin{array}
[c]{c}%
\mathrm{Re}\,\mu>0\\
\mathrm{Re}\,\beta u>0
\end{array}
\end{equation}
and we have momentarily suppressed the suffices $1,2$. The second term of the
integral in $\left(  \ref{Ia}\right)  $ becomes%
\begin{align}
I_{\beta}  &  =\int_{\sqrt{m^{2}}}^{\infty}\exp(-\alpha\frac{E^{2}}{E_{P}^{2}%
})\frac{E^{3}}{E_{P}}\sqrt{E^{2}-m^{2}}dE\nonumber\\
&  \frac{1}{2E_{P}}\int_{\sqrt{m^{2}}}^{\infty}\exp(-\alpha\frac{x}{E_{P}^{2}%
})x\sqrt{x-m^{2}}dx\nonumber\\
&  =\frac{\sqrt{\pi}E_{P}^{4}}{4\alpha^{5/2}}(3+\frac{2\alpha}{E_{P}^{2}}%
m^{2})\exp\left(  -\frac{\alpha m^{2}}{E_{P}^{2}}\right)  ,
\end{align}
where we have used the following relationship%
\begin{equation}
\int_{a}^{\infty}dx\left(  x-a\right)  ^{1/2}x\exp\left(  -\mu x\right)
=\frac{\sqrt{\pi}}{4}\mu^{-5/2}(3+2\mu a)\exp\left(  -\mu a\right)  \qquad
a>0,\mu>0.
\end{equation}

\section{Integrals for Noncommutative Geometry distortion}

\label{app2} In this appendix, we explicitly compute the integrals coming from
$\left(  \ref{t1loop}\right)  $. We begin with%
\begin{align}
&  \int_{\sqrt{m_{0}^{2}\left(  r\right)  }}^{+\infty}\sqrt{\left(  \omega
^{2}-m_{0}^{2}\left(  r\right)  \right)  ^{3}}e^{-\frac{\theta}{4}\left(
\omega^{2}-m_{0}^{2}\left(  r\right)  \right)  }d\omega\label{I1a}\\
&  \underset{\omega^{2}=x}{=}\frac{1}{2}\int_{\sqrt{m_{0}^{2}\left(  r\right)
}}^{+\infty}\sqrt{\left(  x-m_{0}^{2}\left(  r\right)  \right)  ^{3}}%
e^{-\frac{\theta}{4}\left(  x-m_{0}^{2}\left(  r\right)  \right)  }\frac
{dx}{\sqrt{x}}\nonumber\\
&  =\exp\left(  \frac{m_{0}^{2}\left(  r\right)  \theta}{4}\right)  \frac
{1}{2}\left(  \frac{\theta}{4}\right)  ^{-\frac{3}{2}}\sqrt{m_{0}^{2}\left(
r\right)  }\Gamma\left(  \frac{5}{2}\right) \nonumber\\
&  \times\exp\left(  -\frac{m_{0}^{2}\left(  r\right)  \theta}{8}\right)
W_{-1,-1}\left(  \frac{m_{0}^{2}\left(  r\right)  \theta}{4}\right)
,\nonumber
\end{align}
where we have used the following relationship%
\begin{align}
&  \int_{u}^{+\infty}x^{\nu-1}\left(  x-u\right)  ^{\mu-1}e^{-\beta x}dx=\\
&  \beta^{-\frac{\nu+\mu}{2}}u^{\frac{\nu+\mu-2}{2}}\Gamma\left(  \mu\right)
\exp\left(  -\frac{\beta u}{2}\right)  W_{\frac{\nu-\mu}{2},\frac{1-\nu-\mu
}{2}}\left(  \beta u\right) \nonumber\\
&  \operatorname{Re}\mu>0\quad\operatorname{Re}\beta u>0,\nonumber
\end{align}
where $W_{\mu,\nu}\left(  x\right)  $ is the Whittaker function and
$\Gamma\left(  \nu\right)  $ is the gamma function. Further manipulation on
$\left(  \ref{I1a}\right)  $ leads to%
\begin{align}
&  \frac{1}{2}\left(  \frac{\theta}{4}\right)  ^{-2}\left(  \frac{1}%
{2}x\left(  1-x\right)  K_{1}\left(  \frac{x}{2}\right)  +\frac{1}{2}%
x^{2}K_{0}\left(  \frac{x}{2}\right)  \right) \nonumber\\
&  \times\exp\left(  \frac{x}{2}\right)  , \label{I1}%
\end{align}
where%
\begin{equation}
x=\frac{m_{0}^{2}\left(  r\right)  \theta}{4}.
\end{equation}
It is useful to write an asymptotic expansion for $K_{0}\left(  \frac{x}%
{2}\right)  $ and $K_{1}\left(  \frac{x}{2}\right)  $. We get
\begin{equation}%
\begin{array}
[c]{c}%
K_{0}\left(  x/2\right)  \simeq\sqrt{\pi}e^{-x/2}x^{-\frac{1}{2}}\left(
1-\frac{1}{4x}\right)  +O\left(  {x}^{-\frac{5}{2}}\right) \\
K_{1}\left(  x/2\right)  \simeq\sqrt{\pi}e^{-x/2}x^{-\frac{1}{2}}\left(
1+\frac{3}{4x}\right)  +O\left(  {x}^{-\frac{5}{2}}\right)
\end{array}
. \label{Knu}%
\end{equation}
Plugging expansion $\left(  \ref{Knu}\right)  $ into expression $\left(
\ref{I1}\right)  $, one obtains that the asymptotic behavior is given by%
\begin{equation}
\frac{1}{2}\left(  \frac{\theta}{4}\right)  ^{-2}\sqrt{\pi}\left(  \frac{1}%
{2}\sqrt{x}\left(  1-x\right)  \left(  1+\frac{3}{4x}\right)  +\frac{1}%
{2}\sqrt{x^{3}}\left(  1-\frac{1}{4x}\right)  \right)  +O\left(  {x}%
^{-\frac{5}{2}}\right)
\end{equation}
and after a further simplification, one gets%
\begin{equation}
\frac{1}{2}\left(  \frac{\theta}{4}\right)  ^{-2}\frac{3}{8}\sqrt{\frac{\pi
}{x}} \label{asy}%
\end{equation}
while when $x\rightarrow0$, one gets%
\begin{equation}
\frac{1}{2}\left(  \frac{\theta}{4}\right)  ^{-2}\left[  1-\frac{x}{2}+\left(
-\frac{7}{16}-\frac{3}{8}\ln\left(  \frac{x}{4}\right)  -\frac{3}{8}%
\gamma\right)  x^{2}\right]  . \label{ser}%
\end{equation}

\end{document}